# Observation of van Hove Singularities in Twisted Silicene Multilayers

Zhi Li[1], Jincheng Zhuang[1], Lan Chen[2], Yi Du[1,*], Xun Xu[1], Li Wang[1], Xiaodong Pi[3], Xiaolin Wang[1], Kehui Wu[2] and Shi Xue Dou[1]

[1] *Institute for Superconducting and Electronic Materials (ISEM), Australian Institute for Innovative Materials (AIIM), University of Wollongong, Wollongong, NSW 2525, Australia*

[2] *Institute of Physics, Chinese Academy of Sciences, Haidian District, Beijing 100080, China*

[3] *State Key Laboratory of Silicon Materials and Department of Materials Science and Engineering, Zhejiang University, Hangzhou 310027, China*

*To whom correspondence should be addressed. Y. Du: yi_du@uow.edu.au

**Abstract**

Interlayer interactions perturb the electronic structure of two-dimensional materials and lead to new physical phenomena, such as van Hove singularities and Hofstadter's butterfly pattern. Silicene, the recently discovered two-dimensional form of silicon, is quite unique, in that silicon atoms adopt competing $sp^2$ and $sp^3$ hybridization states leading to a low-buckled structure promising relatively strong interlayer interaction. In multilayer silicene, the stacking order provides an important yet rarely explored degree of freedom for tuning its electronic structures through manipulating interlayer coupling. Here, we report the emergence of van Hove singularities in the multilayer silicene created by an interlayer rotation. We demonstrate that even a large-angle rotation ($> 20^\circ$) between stacked silicene layers can generate a Moiré pattern and van Hove singularities due to the strong interlayer coupling in multilayer silicene. Our study suggests an intriguing method for expanding the tunability of the electronic structure for electronic applications in this two-dimensional material.

## Main Text

Low-energy electronic properties of few-layer two-dimensional (2D) materials are known to be strongly dependent on the stacking arrangement [1-3]. Twisted bilayers in Dirac fermion systems, *e.g.* graphene [4], are readily observed, which induce a crossover of Dirac cones that is attributed to rotation of the Brillouin zone (BZ). With interlayer coupling which ensures electron hopping between adjacent layers, the emergence of two saddle points in the band structure due to the overlaid Dirac cones gives rise to logarithmic van Hove singularities (vHs) in the density of states (DOS). When the vHs is close to the Fermi level ($E_F$), its magnified DOS results in electronic instability and consequently, causes new phases of matter with desirable properties, for example, superconductivity, magnetism, and density waves.

Silicene [5-10], a silicon-based Dirac fermion material, has attracted extensive interest since its discovery due to its massless Dirac fermion characteristics [7,10], strong spin-orbital coupling (SOC) [11], and its great potential in electronic applications [8]. The successful fabrication [6,12-14] of silicene subverts the conventional wisdom on hybridization by proving that silicon atoms can form an $sp^2$-$sp^3$ hybridized state and crystallize into a two-dimensional (2D) structure. Although recent scanning tunneling microscopy (STM) and Raman studies have demonstrated that the $sp^3$ component is much relaxed by the low-buckled structure in silicene [7,15], relatively strong interlayer coupling is still expected in multilayer silicene in contrast to the other 2D layered materials, such as graphene and boron nitride. How this strong interlayer interaction perturbs the electronic structure of multilayer silicene, and consequently, leads to new physical phenomena has been rarely studied to date.

All silicene films investigated in this work were grown in a preparation chamber with base pressure ~$5 \times 10^{-11}$ Torr in a commercial low-temperature scanning tunneling microscopy system (LT-STM, Unisoku Co.). Clean Ag(111) substrates were prepared by argon ion sputtering followed by annealing to 820 K for several cycles. The silicene monolayers were then grown on the Ag(111) surfaces by evaporation of silicon atoms from a heated silicon wafer with the substrate kept at 470 K. The detailed growth method has been reported elsewhere [15-18]. The differential conductance, d$I$/d$V$, spectra were acquired by using a standard lock-in technique with a 10 mV modulation at 613 Hz. All the measurements were carried out in ultrahigh vacuum (UHV) at 77 K. To peel off the top silicene layer, we first move the STM tip to a position where we want to peel it off and set sample the bias at -1 V and the tunnelling current at 4 nA to keep a small distance between the STM tip and the silicene surface. Then, we apply a 5 V, 50 ms bias pulse with the feedback loop open with 5 pm proportional gain and 50 m/s integral gain. After this pulse, the top layer of the silicene film will break, and many silicon atoms and clusters will be dispersed on the surface. To reveal what is underneath the silicene surface more clearly, we repeat more pulses. The area of silicene film that one pulse can break depends on the bias of the applied pulse. Bias between 4 V and 5 V is efficient and controllable in our experiments.

To investigate the electronic properties of multilayer silicene, we prepared a sample of multilayer silicene (thickness > 5 monolayers) on an Ag(111) surface by molecular beam epitaxial (MBE). Fig. 1(a) presents an STM image indicating the morphology of the multilayer silicene on the Ag(111) surface. A well-defined √3×√3 honeycomb structure is demonstrated in the top layer of silicene by a high-resolution STM image, as shown in Fig. 1(b). Nevertheless, the √3×√3 honeycomb lattice only consists of the upper buckled silicon atoms, and the detailed structure of the lower silicon atoms is missing. To enhance the resolution of the STM images, we deliberately drove a Pt/Ir tip scanning a step edge of silicene continuously, in order to pick up individual or several silicon atoms on the apex of the tip, as described in previous work [19-21]. As expected, we obtained a high-resolution STM image (Fig. 1(c)) by using Si-terminated tips with a very small tip-sample distance ($V$ = -3 mV, $I$ = 4 nA) for scanning, which indicates a clear 1×1 honeycomb lattice for silicene. In other words, the lower-buckled silicon atoms were also resolved. From the line profile shown in Fig. 1(d), the silicon atoms in the buckled silicene structure have three different apparent heights, which is identical to the predicted ABĀ model in previous work [7,22]. The direct observation of the 1×1 honeycomb lattice of silicene by STM also proves that the √3×√3 phase of silicene is not the Ag-terminated Si(111) reconstruction.

The 1×1 low-buckled structure of the top layer has been determined in our STM investigation, but how can we determine the structure of the underlying silicene layer by STM? Here, we attempted to peel off the top layer of silicene using a bias pulse between the STM tip and the sample. In the peeling-off process, a bias pulse of 5 V with a period of 50 ms was applied on the sample. After several rounds in the same area, the top layer tore and the underlying silicene layer was exposed. Figures 2(a)-2(c) contain STM images of multilayer silicene film before, during, and after the peeling-off process, respectively. We also found that it was much easier to peel off the top layer when there were structural defects nearby, *e.g.* domain boundaries [22], which can weaken interlayer coupling, as shown in Fig. 2(c). The exposed underlying silicene layer still exhibits a √3×√3 structure with the same orientation as the top-layer silicene shown in Fig. 2(d).

In addition to the √3×√3 structure, we also observed a pattern with a large period of about 1.7 nm on the top layer of the silicene film shown in Fig. 3(c), which has never been found in a single layer of silicene. Considering that the √3×√3 lattice is still resolved in STM image [Fig. 3(e)], we conclude that this larger pattern is likely to be a Moiré pattern due to the lattice mismatch between the top layer and the underlying layer of silicene. Furthermore, on comparing the high-resolution STM images obtained from the top layer and the underlying of silicene, we found that the orientations of the two layers were different, with a misorientation angle of 21.8°. If interlayer twisting induces the Moiré pattern, the relationship between the twisted lattices and the Moiré pattern can be described by the Moiré equations: *D*

= $a/[2\sin(\theta/2)]$ and $\phi = 30^{o}-\theta/2$ [23], where *D* is the periodicity of the Moiré pattern, *a* is the lattice constant of √3×√3 silicene, and $\phi$ is the misorientation angle between the √3×√3 silicene lattice and the Moiré superlattice. Taking account of the Moiré periodicity of 1.7 nm and an interlayer rotation angle ($\theta$) of 21.8$^{o}$ in the twisted √3×√3 silicene layers, the angle $\phi$ is about 19.1$^{o}$, which is identical to the measurement on the STM image in Fig. 3(d). Therefore, this angle confirms that the observed Moiré pattern on multilayer silicene originates from interlayer twisting. Interestingly, the protuberance and valley areas in the Moiré pattern can swap their positions under STM scanning, as demonstrated in Figures 3(f)-3(g). The abrupt change in the protuberance and valley areas might result from the dynamic flip-flop motion [22], which is possibly induced by a structural buckling transition in √3×√3 silicene [24].

Exotic electronic properties can be induced by interlayer rotation in 2D layered materials, for instance, van Hove singularities (vHs) in twisted graphene layers and a new set of Dirac fermions in single layer graphene on boron nitride (BN) [1,25]. Fig. 4 shows scanning tunneling spectra (STS) collected on the Moiré pattern in the twisted silicene layers. As shown in Fig. 4(b), two prominent peaks at -1.2 V and 0.75 V can be seen in the STS spectrum taken from the Moiré regions, but are absent in the √3×√3 silicene regions without interlayer twisting. The positions of these two peaks are approximately symmetric with respect to the Dirac point (-0.3 V). We attribute these peaks to interlayer-rotation-induced vHs in the DOS. The mechanism of interlayer-twisting-induced vHs is illustrated in Figures 4(c)-4(d) [22]. In twisted silicene layers, the Dirac cones corresponding to each layer are centered at different points in reciprocal space (indicated as $K_1$ and $K_2$) due to rotation of the Brillouin Zone (BZ) with the same twisting angle ($\theta$). Owing to interlayer electron hopping, the overlapping Dirac cones generate two saddle points that are symmetric with respect to the Dirac point. As a result, they give rise to two vHs peaks in the DOS. Fig. 4(e) is a line profile consisting of 50 spectra collected along the dashed arrow in Fig. 4(a), which shows the spatial distribution of the vHs. It shows that the vHs peaks at 0.75 V and -1.2 V in the twisted silicene can be effectively modulated by the periodic Moiré potential. The asymmetric peak intensity may result from the influence of the third layer of silicene, similar to the case of multilayer graphene [1].

In principle, it is well known that the interlayer coupling parameter, $t_\theta$, depends on the three-dimensional (3D) separation parameter [26] $R = (r^2 + d_\perp^2)^{1/2}$, where $r$ is the spatial separation projected onto the plane, and $d_\perp$ is the interlayer distance. The interlayer coupling is, therefore, modulated by the interlayer rotation, since $R$ increases with increasing interlayer rotation angle ($\theta$). Experimentally, the interlayer coupling strength can be calculated [2] by $t_\theta = (\hbar v_F \Delta K - \Delta E_{vHs})/2$. Here, $v_F$ is the Fermi velocity, $\Delta K$ is the difference between the positions of the Dirac cones in reciprocal space, and $\Delta E_{vHs}$ is the energy difference between the Dirac point and the vHs. Taking twisted graphene as an example, with small interlayer rotation angles ($< 5°$), $t_\theta$ is about 0.108 eV [1,2], and consequently, the vHs is preserved. The interlayer coupling breaks down ($t_\theta \approx 0$ eV) when $\theta$ is greater than 15° in twisted graphene, and thus, the vHs vanishes [27,28]. In the case of twisted silicene, the observed vHs corresponds to a large interlayer rotation angle $\theta = 21.8°$. It is worth noting that 21.8° is the largest rotation angle between two honeycomb lattices that can produce a commensurate superlattice [1]. Considering that $v_F$ is about $5.0 \times 10^5$ ms$^{-1}$ in multilayer silicene, $t_\theta$ is calculated as 0.182 eV in the twisted multilayer silicene sample, which is even greater than it is in twisted graphene with $\theta < 5°$. Hence, it reflects the strong interlayer interaction in multilayer silicene. As observed in STM (Fig. 1), the buckled AB$\overline{\text{A}}$ structure indicates that silicon atoms take $sp^2$-$sp^3$ mixed states in multilayer silicene. The partial $sp^3$ components enhance electron hopping between adjacent layers, in comparison to the case of pure $sp^2$ states, which leads to a stronger interlayer interaction in silicene in contrast to graphene.

In conclusion, we have observed the Moiré pattern and vHs in twisted multilayer silicene. The existence of the 1×1 low-buckled AB$\overline{\text{A}}$ structure has been confirmed. The silicon $sp^2$-$sp^3$ mixed hybridization states lead to a robust interlayer interaction in multilayer silicene, which is much stronger than the interlayer interaction in graphene. It ensures electron hopping between the twisted silicene layers, even with a large interlayer rotation angle. The experimental observations suggest a possible way to engineer electronic properties in multilayer silicene by interlayer twisting.

**Acknowledgements**

This work was supported by the Australian Research Council (ARC) through a Discovery Project (DP 140102581), the Ministry of Science and Technology (MOST) of China (Grant No. 2013CBA01601) and the National Natural Science Foundation (NSF) of China (Grant Nos. 11334011, 11322431). The authors also acknowledge the financial supported by the University of Wollongong through a University Research Council (URC) Small Grant. We gratefully acknowledge discussions with Zheng Liu, Chao Zhang, Zhongshui Ma and Feng Liu. The authors thank Dr T. Silver for valuable support and assistance.

**Figures and figure captions**

**Figure 1**

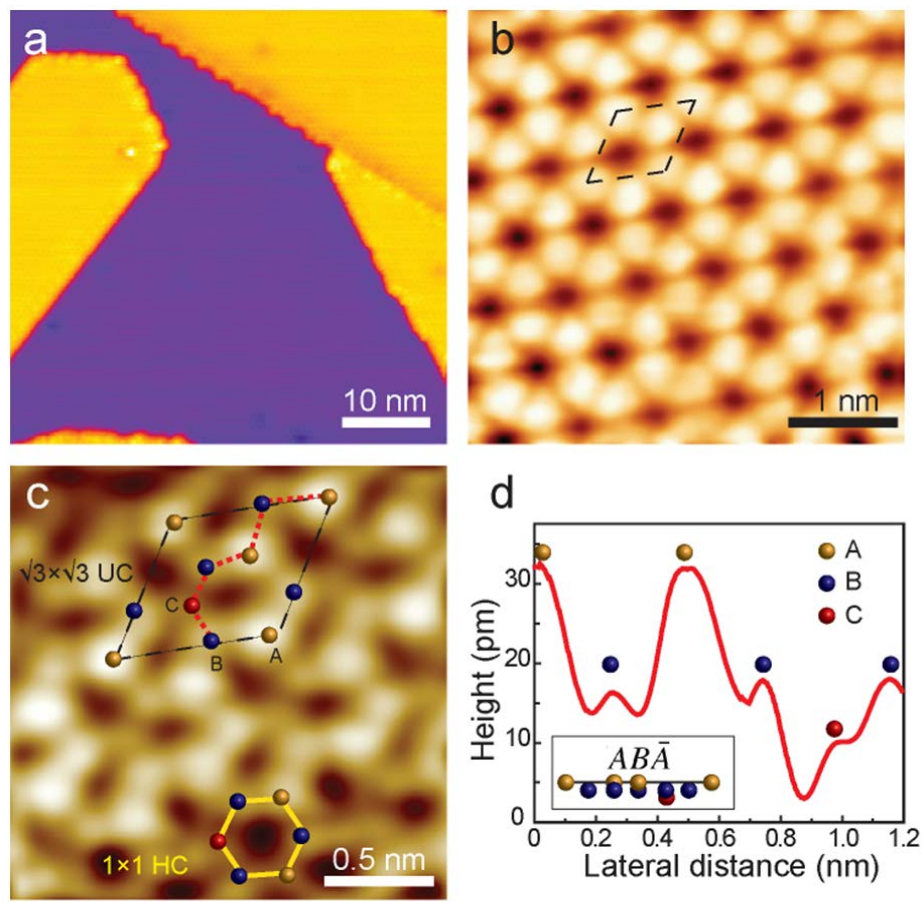

FIG. 1 √3×√3 silicene film and 1×1 honeycomb structure. (a) Typical STM topographic image (50 nm × 50 nm, $V$ = -1 V, $I$ = 5 nA) of the multilayer silicene film. (b) STM image of √3×√3 silicene structure (4 nm × 4 nm, $V$ = -3 mV, $I$ = 4 nA). Only the topmost atoms in the buckled silicene layer are observed, which are arranged in a honeycomb structure. The dashed black rhombus marks a √3×√3 unit cell. (c) High-resolution STM image reveals that the √3×√3 silicene is constructed from the 1×1 honeycomb structures (as indicated by the yellow honeycomb, 1×1 HC). The black dashed line represents the √3×√3 unit cell (√3×√3 UC). The ABĀ buckled structure is reflected by the brightness of the Si atoms. The yellow, blue, and red balls labelled in the √3×√3 silicene unit cell denote the top, middle, and bottom silicon atoms, respectively (2 nm × 2 nm, $V$ = -3 mV, $I$ = 4 nA). (d) Height profile corresponding to the red dashed line in **c**. The inset is a side view of the ABĀ buckled structure corresponding to the height profile.

**Figure 2**

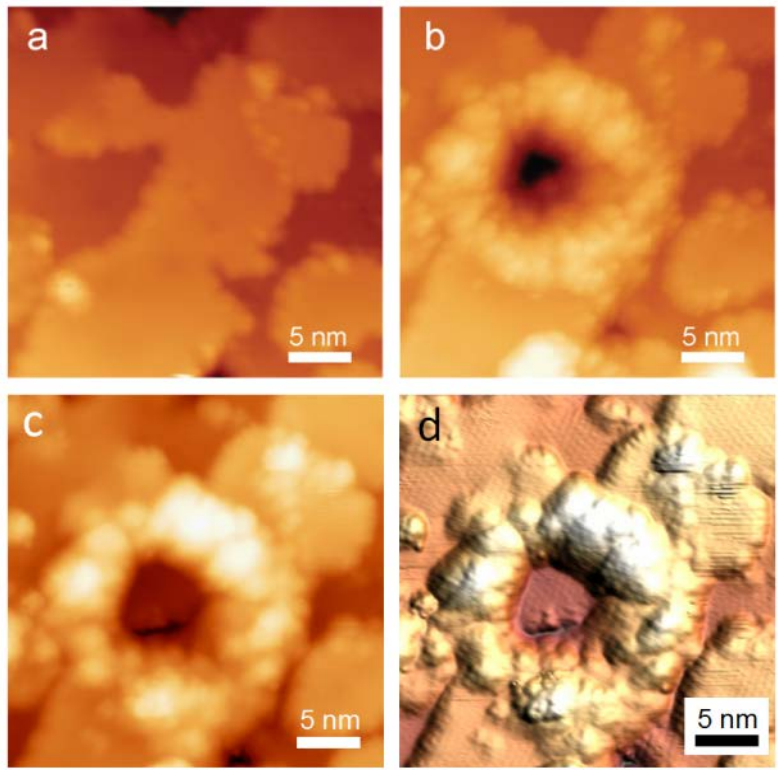

FIG. 2 Peeling-off process for √3×√3 silicene film by using an STM tip. (a) STM image of pristine √3×√3 multilayer silicene. (b) Initial peeling-off process taking place in the same area. The top-layer silicene has been peeled off by the STM tip, leaving a hole-like crack feature. (c) After several rounds of the peeling-off process, the underlying silicene layer is exposed and demonstrates a √3×√3 structure. A domain boundary in the underlying silicene is observed, which possibly weakens the interlayer interaction between adjacent silicene layers. (d) 3D-STM image of **c** shows a clear √3×√3 structure in the underlying silicene layer. (30 nm × 30 nm, $V = -1$ V, $I = 100$ pA).

**Figure 3**

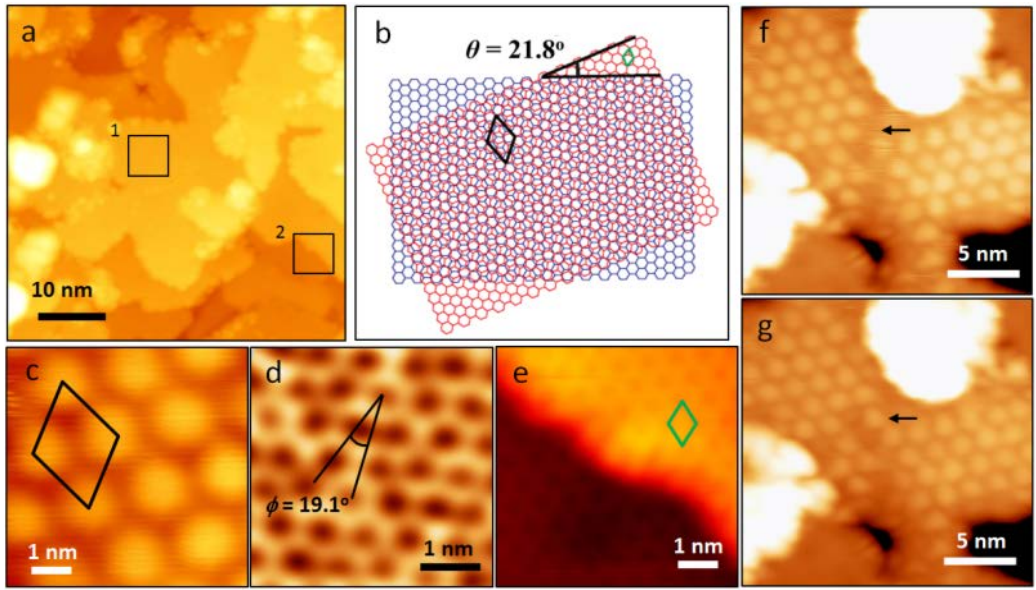

FIG. 3 Moiré pattern in twisted √3×√3 multilayer silicene. (a) Morphology of multilayer silicene (30 nm × 30 nm, $V$ = -1.5 V, $I$ = 100 pA). Regions 1 and 2 indicate a Moiré region and a √3×√3 region, respectively. (b) Schematic diagram of Moiré pattern induced by interlayer twisting in multilayer silicene with a rotation angle ($\theta$) of 21.8°. (c) Enlarged STM image of Region 1 in (a) shows a Moiré superlattice with a lattice constant of 1.7 nm. The Moiré unit cell is outlined by the black rhombus. (6 nm × 6 nm, $V$ = -0.5 V, $I$ = 100 pA). (d) STM image indicates a misorientation angle ($\phi$) of 19.1° between the √3×√3 and the Moiré lattices. (e) Enlarged STM image of Region 2 in (a) demonstrates √3×√3 structure in adjacent layers without interlayer rotation. The unit cell of √3×√3 silicene is outlined in green. (f) and (g), Flip-flop behaviour of the Moiré pattern in a continuous STM scan, which demonstrates the swapping of protuberances and valleys in the Moiré superlattice, as marked by the arrows. (20 nm × 20 nm, $V$ = -0.5 V, $I$ = 100 pA).

**Figure 4**

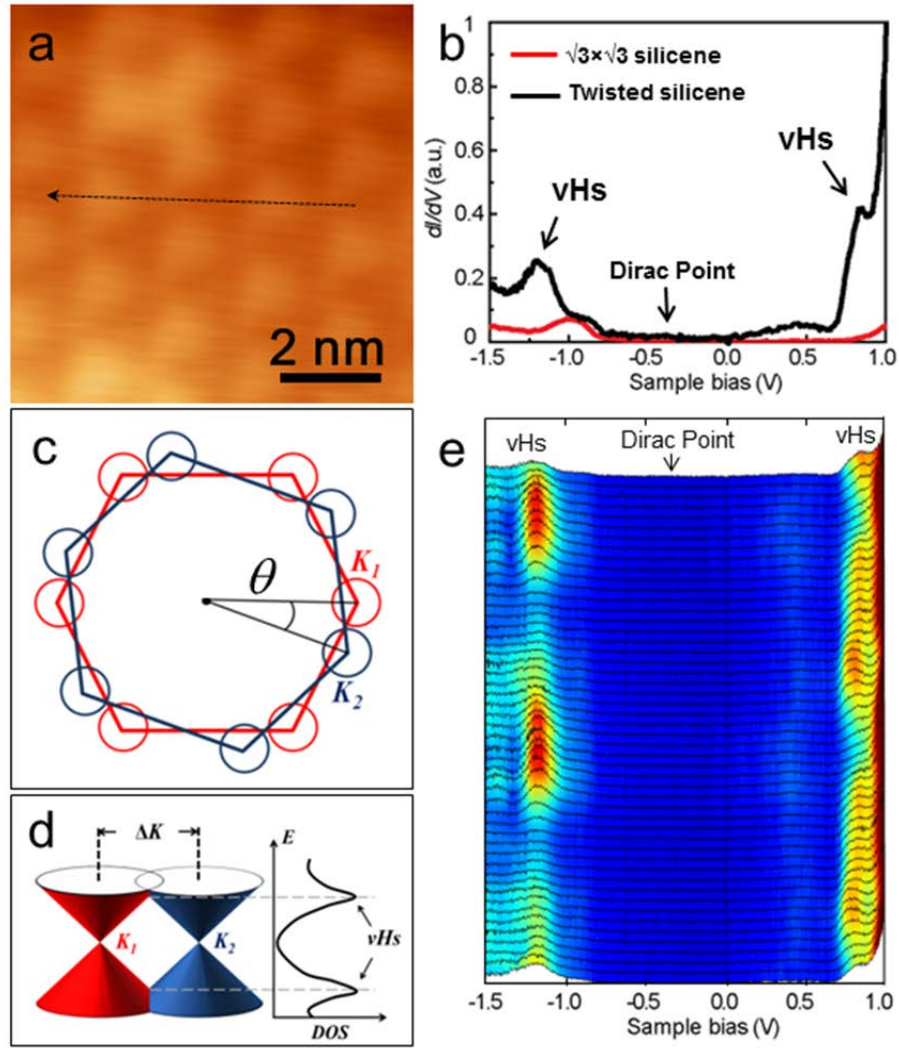

FIG. 4 van Hove singularities in DOS originate from interlayer twisting in multilayer silicene. (a) STM image of Moiré pattern. (b) STS carried out in the Moiré region (Fig. 3(b); set point: $V = 1$ V, $I = 100$ pA) and in the $\sqrt{3}\times\sqrt{3}$ silicene region (Fig. 3(c); set point: $V = 1$ V, $I = 100$ pA). (c) and (d) Illustrations of vHs that originates from twisted silicene layers. The interlayer twisting with the angle $\theta$ leads to a rotation of the BZ with the same angle. The Dirac cones corresponding to $K_1$ and $K_2$ are offset to each other by $\Delta K$. The shifted Dirac cones create saddle points that induce two vHs peaks in the DOS. (e) 50 spectra collected along the black arrow in (a) (from bottom to top; set point: $V = 1$ V, $I = 100$ pA).